\documentclass[a4paper]{article}
\usepackage{etex}
\usepackage{amsmath,amsthm,amstext,amsgen,amsbsy,amsopn,amsfonts,amssymb}
\usepackage{subfigure}
\usepackage{graphics}
\usepackage[pdftex]{graphicx}
\usepackage{epsfig}
\usepackage{bm}
\usepackage{algorithm}
\usepackage{algorithmic}
\usepackage{float}
\usepackage{color}
\usepackage{wrapfig}
\usepackage{esvect}
\usepackage{multirow}
\usepackage{array}

\usepackage{gensymb}
\usepackage{wrapfig}
\usepackage{mathpazo}
\usepackage[normalem]{ulem}
\usepackage{xcolor}
\usepackage{booktabs}

\theoremstyle{plain}
\newtheorem{theorem}{Theorem}[section]
\newtheorem{lemma}[theorem]{Lemma}

\theoremstyle{definition}

\theoremstyle{remark}

\begin{document}


\title{\textbf{A Structured Approach to the Analysis of Remote Sensing Images}}

\author{
Donghui Yan$^{\dag}$, Congcong Li$^{\ddag}$, Na Cong$^{\ddag}$, \\Le Yu$^{\ddag}$, Peng Gong$^{\ddag\P}$
\vspace{0.11in}\\
$^\dag$Mathematics and Data Science, UMass Dartmouth, MA\vspace{0.03in}\\
$^\ddag$Earth System Science, Tsinghua University, China\vspace{0.03in}\\
$^\P$Department of ESPM, University of California Berkeley\\[0.03in]
}

\maketitle

\begin{abstract}
\noindent
The number of studies for the analysis of remote sensing images has been growing exponentially in the
last decades. Many studies, however, only report results---in the form of certain performance metrics---by 
a few selected algorithms on a training and testing sample. While this often yields valuable insights, 
it tells little about some important aspects. For example, one might be interested in understanding 
the nature of a study by the interaction of algorithm, features, and the sample as these collectively 
contribute to the outcome; among these three, which would be a more productive 
direction in improving a study; how to assess the sample quality or the value of a set of features 
etc. With a focus on land-use classification, we advocate the use of a structured analysis. The output 
of a study is viewed as the result of the interplay among three input dimensions: {\it feature, sample, and algorithm}.
Similarly, another dimension, the {\it error}, can be decomposed into error along each input dimension.
Such a structural decomposition of the inputs or error could 
help better understand the nature of the problem and potentially suggest directions for improvement. We 
use the analysis of a remote sensing image at a study site in Guangzhou, China, to demonstrate how such 
a structured analysis could be carried out and what insights it generates. The structured analysis could be
applied to a new study, or as a diagnosis to an existing one. We expect this will inform practice 
in the analysis of remote sensing images, and help advance the state-of-the-art of land-use classification. 
\end{abstract}


\section{Introduction}
\label{section:introduction}
The number of studies for the analysis of remote sensing images has been increasing exponentially in
the last decades. Many studies, however, only report results---in the form of certain performance metrics---by 
a few selected algorithms on a training and testing sample. While this often provides valuable insights 
to practitioners, it tells little about several important aspects. For example, one might be interested 
in understanding a study by the interaction among algorithms, features, and the sample. This is important, 
as these are the factors in a study that involve human decisions which {\it collectively} contribute to the outcome 
of the study. Also of interest is to find out a possible direction for further work in improving 
an existing study---will it be more productive to work really hard on the algorithm, or just focus on finding 
better features, or simply increase the sample size? How much value will it add to increase the sample size? 
This last question arises increasingly often as, after years of practice, the accumulated sample may 
already be fairly large and it is interesting to know if further data collection is worthwhile. 
Additionally, one might be interested in assessing the value of features to decide which features to 
pursue in a future study, or the sample quality to see if the collection procedure needs to be improved.
\\
\\
To shed lights into various important aspects of a study, we advocate the use of a structured analysis. 
We will introduce our approach, particularly, for the land-use classification problem. Our idea was inspired 
by regression diagnosis in statistics \cite{BelsleyKuhWelsch1980,CookWeisberg1983}. {\it Regression 
diagnosis} refers to the assessment of regression analysis, including the validation of various statistical 
assumptions made in regression analysis, the evaluation of variables used in the model, and an 
examination of the influence of individual data points to the model. To better align with the particular 
goals of land-use classification, we re-orient the focus of our structured analysis. While 
regression diagnosis seeks to validate and understand regression results, we aim at a better understanding 
of studies in land-use classification and to identify potential spots for further improvement. 
\\
\\
We take a structured approach. This is to 
overcome the complexity of the land-use classification problem---a number of factors contribute to the outcome and 
some may interact with others in a complicated way. We start by treating the land-use classification as a {\it system} 
with inputs and output. The output is the outcome under some metrics, for example the error rate. The 
inputs are factors that contribute to the outcome, which we identify as three interplaying entities: {\it feature, 
sample, and algorithm}. We term these three entities as the {\it three degrees of freedom} (or dimension) of a study.
Here, {\it feature} refers to the set of features (variables) included in a study, such as vegetation index, quantities
describing the texture pattern in a remote sensing image, values on some spectral bands etc. {\it Sample} are collected 
instances of the tuple, in the form of $(x_1, x_2, ..., x_p, y)$, where $x_i$ is the value 
of the i-th feature, $i=1,2,...,p$, and $y$ is the land-use type. 
{\it Algorithm} is the type of classifiers or models one chooses to use, such as linear models or decision trees etc. 
\\
\\
We view the error as the {\it fourth dimension} of a study. Error can happen to any of the other three dimensions (i.e., 
feature, sample, and algorithm). Distinguishing those can help better understand the study, and to trace the 
contributing source to the outcome. Now that we have identified individual components in a study, how to put those together 
to form a system and to interpret the outcome? That is our {\it structured analysis model}, to be discussed 
in detail in Section~\ref{section:diagModel}. 
\\
\\
A structured analysis will help understand studies in land-use classification. It would yield information that 
connects the dimensions of a study and the observed 
outcome. Such information could help us {\it better understand the results, and potentially suggest directions 
on how to improve the study of land classification}. We will use the analysis of a remote sensing image about 
a study site in Guangzhou, China, to demonstrate how a structured analysis could be carried out. We 
expect this will inform practice in the analysis of remote sensing images, and help advance the state-of-the-art 
of study on the land classification problem. 
\\
\\
It is worthwhile to mention \cite{LiWWHG2014}, a compressive study involving over a dozen of 
different classification algorithms with varying sample sizes. This work gives valuable insights to the practice 
of land classification, including the the importance of sufficient training samples and sample quality etc. In 
contrast, our approach was inspired by regression diagnosis and builds on the theory of pattern classification.
It can be used as a general framework for the analysis of a particular study site, or to understand, evaluate, 
or improve various aspects of an existing analysis (thus could be viewed as a meta analysis). Our approach 
considers all important aspects in a land-use classification analysis, including their interactions and tradeoff 
etc, and gives methodological guideline to practice. While there are common elements with \cite{LiWWHG2014} 
on the assessment of training samples and algorithms, we delve further and towards broader issues. Our 
approach helps to decide if the training samples are sufficient, if the algorithms used are rich enough to capture 
the patterns in the particular land-use problem, how to assess or compare the importance of features, what are the 
difficult land-use types to classify, what are the possible, or most profitable, directions (among sample, feature, 
or algorithm) to work on for further improvement of a study etc.  

\section{A model for structured analysis}
\label{section:diagModel}
\begin{figure}[h]
\centering
\begin{center}
\hspace{0cm}
\includegraphics[scale=0.38,clip]{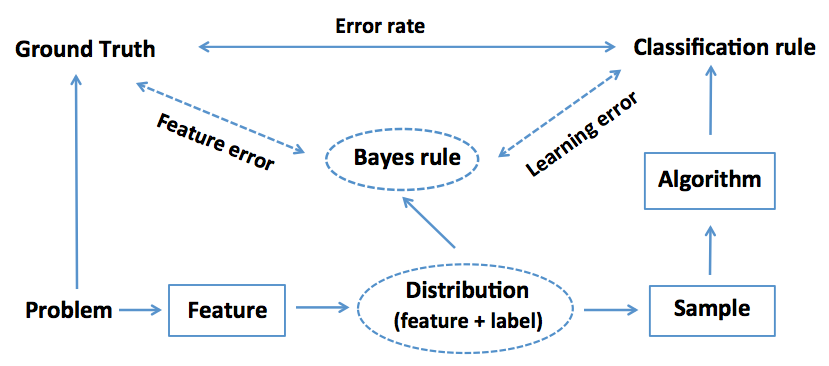}
\end{center}
\abovecaptionskip=-3pt
\caption{\it A model for structured analysis. } 
\label{figure:diagModel}
\end{figure}

In this section, we will introduce our model for structured analysis. To make the model more interpretable, 
we include {\it ground truth}, and two additional `virtual' entities: {\it the probability distribution and the Bayes 
rule}. By `virtual' we mean entities not observable, but are fundamental in land-use classification. 
Figure~\ref{figure:diagModel} is an illustration of our model. Note that the three rectangles indicate entities 
that involve one's choices and decisions, while entities enclosed by a dashed oval are virtual entities. For 
the rest of this section, we will explain individual entities in the model. 
\\
\\
The {\it probability distribution} tells how the values of the (feature, label) pair, denoted by $({\bm{X}},Y)$, look 
like in the data space. It is decided by the features (i.e., variables) used in the study and the nature of the given land-use 
classification problem. The distribution determines the actual classification problem we work with, and, consequently, 
the lowest possible error rate achievable by any classifiers, i.e., the {\it Bayes rate}. The classifier that achieves 
the Bayes rate is called a {\it Bayes rule}. Once the set of features is chosen, the Bayes rate is the theoretical lowest 
possible error rate one can achieve, regardless of how hard one works on improving the classification algorithm 
or how big the training sample is. 
\\
\\
For every land-use classification problem, there is a {\it ground truth}, which always tells the correct label.
When one chooses to use a particular set of features in a study, 
there is often a loss of information (since there are other features potentially informative but not used).
This would cause a gap between the Bayes rule and ground truth. We call this {\it feature error}. 
To reduce the gap, one needs to improve on feature selection. 
\\
\\
The idea of classification is to find a mapping $f$ between the feature $\bm{X}$ and the label $Y$. 
This requires knowledge about the probability distribution, which is generally unknown; what we have 
is a {\it sample} collected from this distribution. We wish to use the sample to estimate the mapping $f$; 
the estimated mapping $\hat{f}$ is called a {\it classification rule}. The sample size can be changed, 
depending on the availability. Often, a large sample is desired. However, after certain point, 
the gain in performance diminishes when further increasing the sample size. 
\\
\\
Now given a collected sample, we need an algorithm to fit a {\it classification rule} (i.e., to find the 
estimated mapping $\hat{f}$). By {\it algorithm} 
we mean the type of classifiers or models, such as linear models or decision trees etc, used to fit 
the classification rule. Different choices of algorithms lead to different types of classification rules. 
The fitted classification rule will be used for classification on the test sample. With reference to the 
ground truth, one can calculate the {\it error rate}, which is the proportion of the test sample that 
receives a wrong label. 
\subsection{The errors in the structured analysis model}
\label{section:errModel}
The errors play an important role in our model of structured analysis. While the classification error rate 
measures the final outcome, it is a little crude. It will be helpful to decompose the classification error
according to the error sources. There are three sources of errors, corresponding to {\it feature error, 
sample error, and learning error}, respectively. We have discussed the feature error, next we will 
discuss sample error and learning error. 
\\
\\
The {\it learning error} results from the training of the classifier. In practice, we know neither $f$ nor the 
probability distribution. We wish to use a training sample collected from the unknown distribution
to learn the classification rule with some algorithm. There are two potential errors. One 
is the {\it approximation error} due to the inappropriate choice of the type of algorithms. For example, 
for a particular problem, a boosting \cite{AdaBoost} type of algorithms work the best but support vector 
machine \cite{SVM} or a simple linear model is used. Another is called the {\it convergence error}, due to the insufficient size of 
the training sample. One could lower the convergence error by increasing the sample size, while the
approximate error could be reduced by increasing the richness of the family of classification rules in model 
fitting (one could try different algorithms when there is not much information about the problem structure).  
\\
\\
The {\it sample error} refers to the discrepancy between the true probability distribution of $({\bm{X}},Y)$ 
and that of the collected sample. It is related to the data quality or whether the sample is representative of 
the true probability distribution. The representativeness of the sample is related to the study design. Usually 
the principle of random sampling \cite{Rice1995} is followed. There are generally two types of errors related 
to data quality, namely, {\it data perturbation} \cite{ZhuWu2004, HuangYanNIPS2008, NguyenHuang2008} 
and {\it data contamination} \cite{Tukey1960, ZhuWu2004, cdc}. Data perturbation is often caused by additive 
noise and would affect a large proportion of the data, typically at a small amount. Data contamination substitutes
a random subset of the data by a different distribution. Both will impact the accuracy of the land classification.
\\
\\
On an orthogonal direction, one may decompose the error according to the land types. Which land-use 
types are frequently misclassified? Or misclassified into which land-use types? This could be done with a confusion 
matrix to be discussed in Section~\ref{section:confMatrix}. Such information would become
useful clues in the search of better algorithms or new features.
\section{Study site and the data}
\label{section:siteData}
\begin{figure}[h]
\centering
\begin{center}
\hspace{0cm}
\includegraphics[scale=0.36,clip]{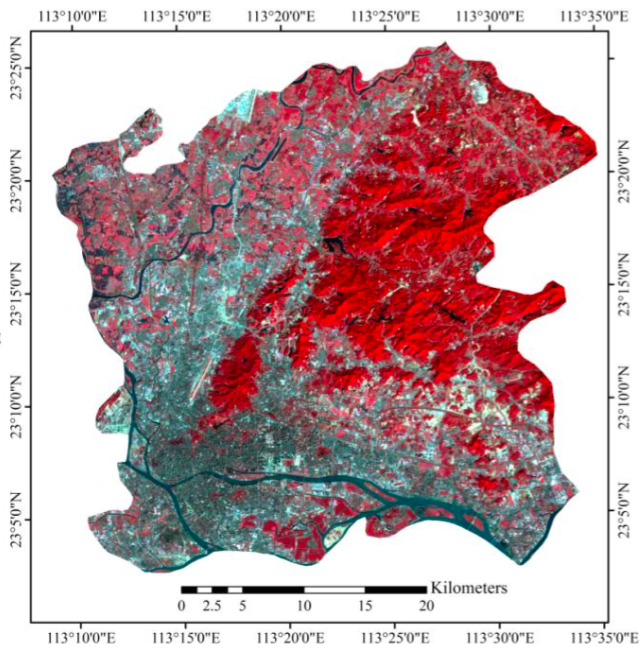}
\end{center}
\abovecaptionskip=-5pt
\caption{\it The study area. The image displays the green, red and near infrared band of the TM data with blue, green, and red colour guns.} 
\label{figure:studyArea}
\end{figure}

Our study site is located in the Pearl River Delta, or more specifically, the region spanning 23\degree 2'-23\degree 25'N, 113\degree 8'-113\degree 35'E, in Guangdong Province of South China. The study site contains the central part of Guangzhou and its rural-urban fringe. Figure~\ref{figure:studyArea} is a Landsat Thematic Mapper (TM) image for the study site. As Guangzhou has undergone rapid urban development in the last two decades, it has been studied extensively for land use, land cover mapping and change detection; see, for example, \cite{SetoWSHLK2002,FanWengWang2007,FanWQW2009,LiWWHG2014}.
The Landsat TM image for the study site was acquired on 2 January 2009, in the dry season of this area. The raw imagery was geo-referenced in 2005 with a root mean squared error of 0.44 pixels. A 6-band set of the TM data was used (excluding the thermal band due to its coarse resolution). 
\\
\\
With reference to some popular land cover and land-use classification systems \cite{GongHowarth1992SPOT,GongMarceau1992, GongMapping2013,LiWWHG2014}, 
7 different land-use types (a.k.a. classes) are used in our study. A brief description of the land-use types is given in Table~\ref{table:landType}. 
\begin{table}[h]
\begin{center}
\begin{tabular}{rr|l}    
\toprule
&\textbf{Land-use type} & \textbf{Description}\\[1pt]
\midrule
&{\it Water}  & Water bodies such as reservoirs, ponds and river\\
&{\it Residential area}  & Areas where driveways and roof tops dominate \\
& {\it Natural forest} & Large area of trees \\
&{\it Orchard}       & Large area of fruit trees \\
& {\it Industrial/commercial}    & Lands where roof tops of large buildings dominate\\
&{\it Idle land}      & Lands where no vigorous vegetation grows\\
& {\it Bareland}  & Lands where vegetation is denuded or where the\\
&                        & construction is underway\\ 
\bottomrule
\end{tabular}
\end{center}
\caption{\it A description of 7 different land-use types.}
\label{table:landType}
\end{table}
The training and test samples are adopted from a recent study \cite{LiWWHG2014}. The training sample size is 2880, and the number of instances are $240,240,480,240,960,240,480$, respectively, for the 7 land-use types in the order listed in Table~\ref{table:landType}. The test sample has a size of 423, 
with a class distribution of 
\begin{equation*}
(9.69\%, 21.51\%, 19.62\%, 11.35\%, 16.78\%, 10.64\%, 10.40\%).
\end{equation*}
We use the classification error as the evaluation metric as this is common in the data mining and also the remote sensing literature (note another popular 
metric is the Kappa statistic); also we will use a quantity, the {\it distance of separation} 
to be discussed in Section~\ref{section:theory} to assess the relative strength of different features.    
\\
\\
We use a total of 56 features. There are 6 spectral features corresponding to the 6 TM bands,
including {\it blue, green, red, near infrared, shortwave infrared 1}, and {\it shortwave infrared 2}, respectively. 
Each TM band corresponds to 8 texture features, including {\it mean, variance, homogeneity, 
contrast, dissimilarity, entropy, second moment, correlation}; this gives a total of 48 texture features. 
Additionally, there are two location features, the {\it latitude} and {\it longitude} of the ground position 
associated with each data instance. Table~\ref{table:features} is a summary of the features.
\begin{table}[h]
\begin{center}
\begin{tabular}{rr|l}    \toprule
&\textbf{Feature code}                                 & \textbf{Description}  \\
    \midrule
&{\it Lat, Lon}                                   & Latitude, longitude \\
&{\it B1, B2, ..., B6}                                        & Spectral features for the 6 TM bands\\
&{\it B7, B8, ..., B54}                                      & Texture features. Each of the 6 TM bands\\
&                                                                    & corresponds to 8 texture features \\ \bottomrule
\end{tabular}
\end{center}
\caption{\it A list of all features used (totally 56). The texture features for each of the 6 bands are listed in the 
order of mean, variance, homogeneity, contrast, dissimilarity, entropy, second moment, and correlation.}
\label{table:features}
\end{table}
\section{Tools for structured analysis}
\label{section:theory}
In a land-use classification study, we are often interested in several important questions. How 
good is a particular 
set of features? What might be the contribution of individual features? Will it add value, or 
how much value would it be, by adding another set of new features? What would one expect 
on the predictive accuracy from the `best' algorithm if he has `enough' computing power and 
sample? Which land-use types are more prune to classification errors? To gain insights into 
these questions, we propose to study several quantities, including the {\it covariance matrix} 
of the features, the confusion matrix of errors, and the {\it distance of separation} of the data 
(under a given set of features). For the rest of this section, we will introduce these along with 
a characterization on when combining two sets of features may be beneficial. 
\subsection{The covariance matrix}
For a given set of features, a central quantity in characterizing the data distribution is the 
covariance structure of the features. This is described by the covariance matrix, denoted by 
$\bm{\Sigma}$, with its $(i,j)$-position defining the covariance between the $i^{th}$ and $j^{th}$ 
feature. That is, 
\begin{equation}
\bm{\Sigma}_{i,j} = \mathbb{E} (X^{i} - {\mu}^i)(X^{j} - \mu^j),
\end{equation}
where $\mathbb{E}$ indicates expectation, $\mu^{i}$ and $\mu^{j}$ are the mean of the $i^{th}$ and $j^{th}$ 
feature which are denoted by $X^{i}$ and $X^{j}$, respectively.
In practice, one often scale each feature to have a variance 1 and this leads to the correlation matrix. 
To abuse the notation a bit, we still use $\bm{\Sigma}$ for the correlation matrix.
All entries of the correlation matrix are in the range $[-1,1]$. A small $|\bm{\Sigma}_{i,j}|$ indicates
a low correlation between the $i^{th}$ and $j^{th}$ feature; otherwise there would be a collinearity 
among features and special cares (e.g., regularization) are needed in model fitting. If the features
jointly follow a normal distribution, then $|\bm{\Sigma}_{i,j}|=0$ is equivalent to the independence of the 
two features.
\subsection{The confusion matrix}
\label{section:confMatrix}
The confusion matrix \cite{Ting2011} is a two-way table that summarizes the test instances according 
to their actual class and predicted class. It has the following form:
\small
\begin{center}
{\begin{tabular}{c|ccccc|c}
      & 1        &  ...    &  j       &  ... & C        & Total \\[2pt]
    \hline
1     & $n_{11}$ &  ...    & $n_{1j}$ &  ... & $n_{1C}$ & $n_{1.}$ \\
...   & ...      &  ...    & ...      &  ... &  ...     & ...      \\
i     & $n_{i1}$ &  ...    & $n_{ij}$ &  ... & $n_{iC}$ & $n_{i.}$ \\
...   & ...      &  ...    & ...      &  ... & ...      & ...      \\
C     & $n_{C1}$ &  ...    & $n_{Cj}$ &  ... & $n_{CC}$ & $n_{C.}$ \\[3pt]
\hline\\[-8pt]
Total & $n_{.1}$ &  ...    & $n_{.j}$ &  ... & $n_{.C}$ & n        \\
\end{tabular}}
\end{center}
\normalsize
\bigskip
where the columns indicate the true land-use types (classes) and the rows predicted ones, C is the
number of different classes, $n_{ij}$ is the number of instances from class $i$ but 
classified as being from class $j$, $n_{i.}$'s are the row sums and $n_{.j}$'s are the column
sums of the table, and $n$ is the size of the test sample. The numbers on the diagonal are the instances correctly
classified while off-diagonals are misclassified. The confusion matrix allows one to see where the errors 
are by classes. This will help narrow down the focus to a few hard to classify land-use types, and suggest 
directions for further study.
\subsection{The distance of separation}
\begin{figure}[h]
\centering
\begin{center}
\hspace{0cm}
\includegraphics[scale=0.36,clip]{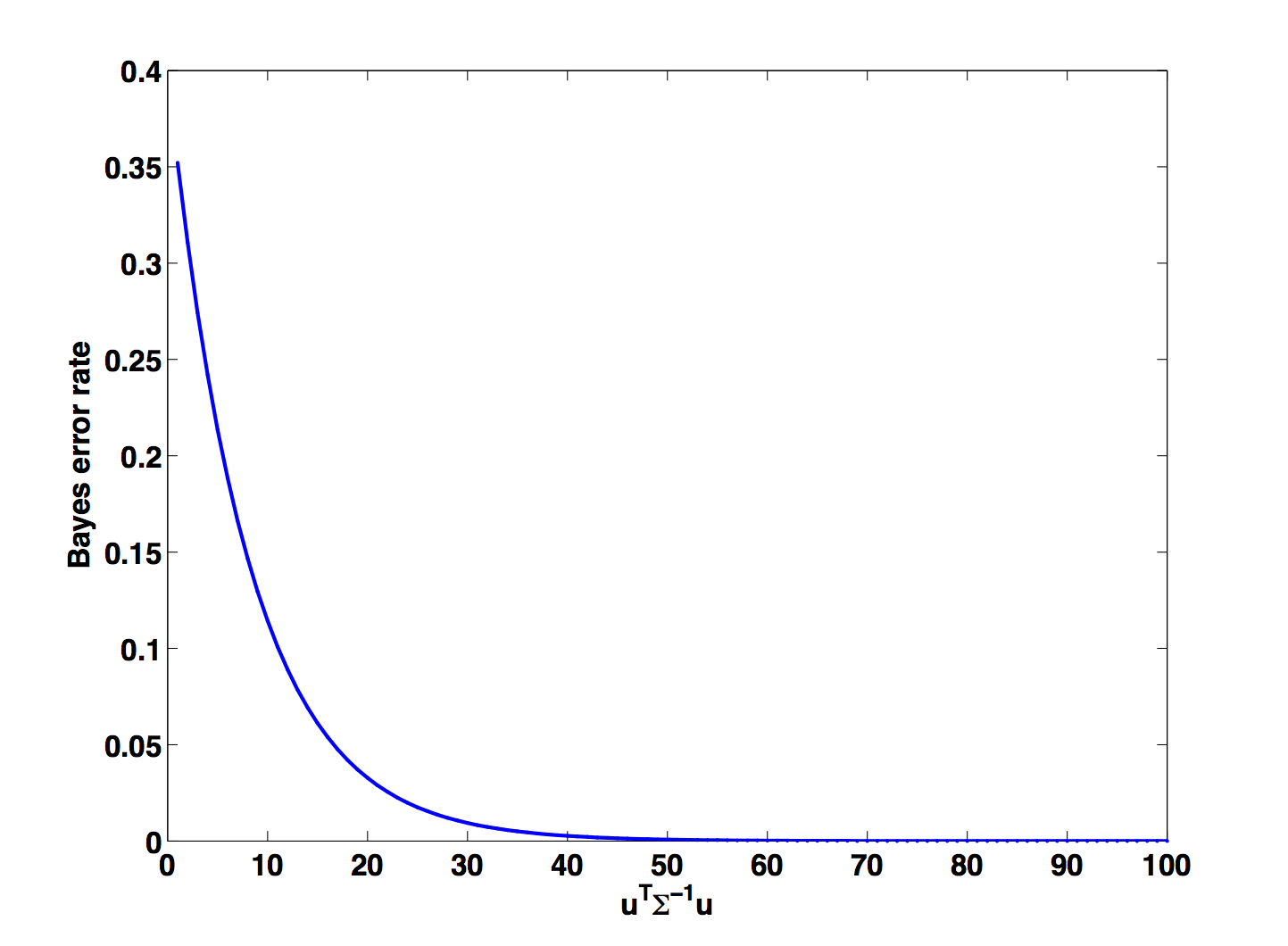}
\end{center}
\abovecaptionskip=-5pt
\caption{\it Bayes error rate as a function of the distance of separation.} 
\label{figure:BayesErrorSeparation}
\end{figure}

The {\it distance of separation} was studied by \cite{TACOMA} as an indication of the strength 
of a set of features. 
The associated theoretical model is the Gaussian mixture, due to it versatility 
in modeling the real data \cite{MclachlanPeel2000}. For simplicity, we consider the 2-component 
Gaussian mixture specified as
\begin{eqnarray}
\label{eq:gm} \Theta\mathcal{N}(\bm{\mu},\bm{\Sigma}) +(1-\Theta)
\mathcal{N}(-\bm{\mu},\bm{\Sigma}),
\end{eqnarray}
where $\Theta \in \{0,1\}$ indicates the label of an observation such
that $\mathbb{P}(\Theta=1)=\theta$, and $\mathcal{N}(\bm{\mu},\bm{\Sigma})$
stands for Gaussian distribution with mean $\bm{\mu} \in
\mathbb{R}^p$ and covariance matrix $\bm{\Sigma}$. Here W.L.O.G., we assume 
the center of the mixture components are $\pm \bm{\mu}$. This can be achieved
by shifting the data without changing the nature of the problem. For simplicity, 
we consider $\theta=\frac{1}{2}$ and the 0-1 loss.
\\
\\
The {\it distance of separation} is defined as
\begin{equation}
d_{\bm{\mathcal{F}}}=\bm{\mu}_{\bm{\mathcal{F}}}^T\bm{\Sigma}^{-1}\bm{\mu}_{\bm{\mathcal{F}}},
\label{eq:distSep}
\end{equation}
where $\bm{\mathcal{F}}$ indicates a set of features, $\bm{\Sigma}$ and $\pm \bm{\mu}$ are as
defined in \eqref{eq:gm}.
At an intuitive level, one can view $d_{\bm{\mathcal{F}}}$ as indicating how far apart 
the data is between different classes---the larger this distance is, the data are further
apart thus easier for a classification algorithm to locate the class boundary. It is related 
to the Bayes error of classification for which there is a well-known result.
\begin{lemma}[\cite{Anderson1958,DudaHart1973}]
\label{lemma:lemmaAnderson} For Gaussian mixture \eqref{eq:gm} and
0-1 loss, the Bayes error rate is given by
$\Phi(-\frac{1}{2}(\bm{\mu}_{\bm{\mathcal{F}}}^T\bm{\Sigma}^{-1}\bm{\mu}_{\bm{\mathcal{F}}})^{1/2})$
where $\Phi(\cdot)$ is defined as $=\int_{-\infty}^{x}
\frac{1}{\sqrt{2\pi}}e^{-\frac{z^2}{2}}dz$.
\end{lemma}
\noindent
To better appreciate the role played by the distance of separation in Bayes error, we plot  in
Figure~\ref{figure:BayesErrorSeparation} the Bayes error as a function of the distance of separation. 
It can be seen that the Bayes error decreases exponentially fast as the distance of separation increases. 
\\
\\
The connection between the distance of separation and Bayes error allows us to quantify the 
`strength' of a feature set. The larger the distance of separation, the smaller the 
Bayes error (by Lemma~\ref{lemma:lemmaAnderson}), and consequently, the smaller the
feature error (as the {\it ground truth} is always correct) by Figure~\ref{figure:diagModel}. 
It should be noted, however, that to translate the strength of a feature set to empirical 
performance, the training sample size needs to grow proportionally and the classifier 
is rich enough to match the complexity of the problem. 
\\
\\
Additionally, when using the empirical distance of separation, one should keep in mind that 
such an estimate would only serve the purpose of giving a qualitative characterization rather 
than quantifying the actual Bayes error. The is because it involves the estimation of the 
covariance matrix which is notoriously difficult when the number of features is large 
\cite{BickelLevina2008b,FanLiaoLiu2016}. 
\subsection{The marginal benefit}
One can also use the distance of separation to study the {\it marginal benefit} of one set of features w.r.t. 
another. If {\it new} features cause the distance of separation to increase thus a smaller Bayes 
error, then it is {\it beneficial} to add such new features\footnote{When the set of new features 
are not noises, it always increases the distance of separation. For practical reason, it only helps
when such an increase is substantial. The estimated distance of separation allows us to see 
whether this is true.}. Again, a sufficiently large training sample size is required for the reduced 
Bayes error to materialize; otherwise, it may be harmful to the empirical performance due to a 
potential overfit caused by the small sample size. 
\\
\\
In this section, we will characterize a situation where the inclusion of a set of new features 
will be marginally beneficial. Roughly, we require the set of new features to posses 
discriminative power and that the two sets of features have `low' dependence. The discriminative power is 
equivalent to a positive distance of separation, as a 0 distance of separation would result in a 
random guess, i.e., 50\% Bayes error for a two-class classification problem. 
\\
\\
Let the covariance matrix $\bm{\Sigma}$ be written as
\[\bm{\Sigma} = \left[
\begin{array}{cc}
\bm{A}_{11}              & \bm{A}_{12}\\
\bm{A}_{12}^T             & \bm{A}_{22}\\
\end{array}
\right]\] where we assume block $\bm{A}_{11}$ and $\bm{A}_{22}$ correspond to two sets of features, respectively, 
after a permutation of rows and columns of $\bm{\Sigma}$. Correspondingly, write 
$\bm{u}_{\bm{\mathcal{F}}}=[\bm{u}_1, \bm{u}_2]^T$ and $\bm{\mathcal{F}}=[\bm{\mathcal{F}}_1, \bm{\mathcal{F}}_2]^T$. 
We assume $||\bm{A}_{12}||_{F}=o(1)$ for a `low' dependence between two sets of features
$\bm{\mathcal{F}}_1$ and $\bm{\mathcal{F}}_2$; here $||.||_{F}$ 
denotes the Frobenius norm \cite{GolubVanLoan1989} and $o(1)$ is the little-o notation indicating
that the quantity is small compared to 1. Our main result can be stated as the following theorem.
\begin{theorem}
\label{thm:largerSeparation} Suppose the data is generated according to \eqref{eq:gm}. Assume 
$\bm{u}_2^T \bm{A}_{22}^{-1} \bm{u}_2>c>0$, $||\bm{A}_{12}||_{F}=o(1)$, $||\bm{u}||_F \leq C$ for some
positive constants $c$ and $C$, and that the eigenvalues of both $\bm{A}_{11}$
and $\bm{A}_{22}$ are bounded away from $0$ and $\infty$. Then
\begin{equation}
\label{eq:largerSeparation}
\bm{u}_{\bm{\mathcal{F}}}^T\bm{\Sigma}^{-1}\bm{u}_{\bm{\mathcal{F}}}
> \bm{u}_1^T \bm{A}_{11}^{-1} \bm{u}_1.
\end{equation}
\end{theorem}
\noindent
The proof of Theorem~\ref{thm:largerSeparation} follows a similar line of arguments as \cite{TACOMA}, 
and is given in the appendix. Rather than discussing the technical details, we will give here a few remarks 
on the interpretation and implication of Theorem~\ref{thm:largerSeparation}. 
\\
\\
\textbf{Remarks.} 1). It is beneficial to combine two sets of features with low correlation (provided that the 
training sample is sufficiently large and the family of classifiers is rich enough). The theorem states 
that this would lead to a larger distance of separation thus a decreased Bayes error. 
\\2). Setting $||\bm{A}_{12}||_{F}=0$ recovers the independence case. So the independence 
case is a special case of Theorem~\ref{thm:largerSeparation}.
\\3). Extra features will not help much if the existing features are already good enough, 
i.e., $\bm{u}_1^T \bm{A}_{11}^{-1} \bm{u}_1$ is big. In such a case, the Bayes error 
under the existing features is already very small, and there is not much room for improvement.
\section{Methods}
\label{section:methods}
Our implementation of a structured analysis focus on the four dimensions of a study, 
including sample, algorithm, feature, and the error, as well as a decomposition of the 
error by the confusion matrix. We will discuss each in this section.
\subsection{Sample}
\label{section:methodData}
The sample is an important dimension in a study. While we have decoupled the {\it feature} aspect 
from the data (c.f. Section~\ref{section:diagModel}), there remain several aspects of importance (assuming 
the data has been properly cleaned and pre-processed). These include the {\it size}, {\it quality}, 
and {\it representativeness} of the data. 
\\
\\
The size of the sample is related to the {\it convergence error} in model fitting. Typically, a larger sample 
would improve the predictive accuracy, but often that is not feasible in practice. Also, one may wish to know how 
much improvement to expect when there is a larger sample. We propose to {\it subsample} the training 
set at varying sizes to see the trend of the error rate vs the training sample size. This will help 
probe the convergence error, and to see if a larger sample will likely lead to a notable improvement in the 
performance. 
\\
\\
Additionally, when inspecting the confusion matrix, we suggest subsample the training 
set to extrapolate how the confusion matrix will change when the sample size increases. Of particular interests 
are those cells, or rows, in the confusion matrix indicating a substantially higher error rate than others. 
These will allow us to focus on those challenging land-types, and we can then examine if the algorithm or the 
features are adequate for those land-types.
\\
\\
To assess the data quality or representativeness of the sample is a hard problem, as the true probability 
distribution is unknown. For the particular sample used in this study, it is possible to carry out hypothesis 
testing according to the ground position (latitude and longitude) associated with a pixel. That is, to test 
whether the set of (latitude, longitude) pairs from the sample have a uniform distribution over the study region. 
However, that would require the (latitude, longitude) information, and that the study region has a regular 
shape (so that it would be easy to do computation), which are typically not applicable in other studies, we 
omit the discussion here.
\\
\\
For land classification problems, feature noises are mainly caused by noises to the remote sensing images. 
Fortunately, the advance in remote sensing technology has now made it much less of a concern than the 
label noise. So here we focus on the label noise and its potential impact. It is in general hard to 
estimate the amount of label noise, we suggest the following procedure to probe it. Randomly select a 
proportion, $\epsilon \in [0,1]$, of data and then flip the labels uniformly at random to a different 
label. The prediction error on the clean (uncontaminated) test set for each $\epsilon$ form a curve of test errors 
vs $\epsilon$. This curve allows us to extrapolate the amount of label noise in the original sample or its 
impact. $\epsilon$ is estimated to be smaller than 10\% in many applications according to \cite{Hampel1974}. 
We recommend trying several different classification algorithms, particularly Random Forests (RF, 
\cite{RF}) which has a reputation of strong noise resistance. If the curves 
by different classifiers are all steep, that is an indication of potentially non-negligible label noise; if at least 
one curve is relatively flat, then either the label noise is small or its impact can be safely ignored. 
\subsection{Algorithms}
\label{section:methodAlgorithms}
The algorithm is another dimension of a study. It is related mainly to the approximation error in model fitting.
The richness of the family of classifiers is required to `match' the complexity of the classification problem in 
order to have a small approximation error. The complexity of the problem is determined by the distribution of 
$({\bm{X}},Y)$, which is often unknown. To probe this, we recommend trying several different types of algorithms, 
hoping that some would have a matching richness. A number of existing studies \cite{metaLandsat2014, LiWWHG2014} 
are actually along this line. Of course, different types of algorithms may have a different convergence rate (faster 
convergence implies a smaller convergence error for a given sample size). Here convergence indicates that the 
classification algorithm has reached a state that further increasing the sampling size will no longer cause much 
changes to the classification rule. When the sample size is `small', it is highly desirable to explore a range of 
different types of algorithms.  
\\
\\
Since many different algorithms have already been explored in \cite{LiWWHG2014}, we choose to use
two of best performing ones, RF and $L_1$-regularized logistic 
regression \cite{glmnet2010}. RF is widely acknowledged as one of the most powerful tools in statistics and 
machine learning according to some empirical studies \cite{RF,Caruana2006, caruanaKY2008}. 
Regularized logistic regression is a popular algorithm that combines a superior 
predictive performance with a strong variable selection capability. 
\\
\\
RF is an ensemble of decision trees. Each tree is built by recursively partitioning 
the data. At each node (the root node corresponds to the bootstrap sample), RF randomly 
samples (with replacement) a number of features and then select one for 
an `optimal' partition of that node. This process continues recursively until the 
tree is fully grown, that is, only one data point is left at each leaf node. RF often has superior 
empirical performance, is very easy to use (e.g., very few tuning parameters) and show a 
remarkable built-in ability for feature selection. We will use the R package  {\it randomForest}.  
\\
\\
Logistic regression models the log odds ratio of the posterior probability as a linear function 
of the co-variates (i.e., features):
\begin{eqnarray}
&& log\frac{P(Y=1 | \bm{X}=\bm{x})}{1-P(Y=1 | \bm{X}=\bm{x})} = \sum_{i=1}^p \beta_i x_i,
\label{eq:logit}
\end{eqnarray}
where $\bm{x}=(x_1,...,x_p)$.
When there is a potential high collinearity among the features, and, especially when the 
number of features is large w.r.t. the sample size, typically regularization is used. {\it Regularization} 
\cite{Tikhonov1963} is the idea of injecting external knowledge, e.g., smoothness 
\cite{KimeldorfWahba1970,Wahba1990} or sparsity \cite{ChenDonoho1994, DonohoJohnstone1994,Tibshirani1996} 
etc, into model fitting. A popular form of regularization is to enforce an $L_1$-penalty 
on the coefficients \cite{Tibshirani1996,ParkHastie2007,glmnet2010}. This leads to the following 
$L_1$-regularized logistic regression:
\begin{eqnarray}
&& \arg \max_{\beta=(\beta_1,...\beta_p)} L(\beta)=\Pi_{i=1}^n \left[ p(\bm{X}_i) \right]^{Y_i} \left[ 1-p(\bm{X}_i) \right]^{1-Y_i} 
+ \lambda \sum_{i=1}^p |\beta_i|, 
\label{eq:logitL1}
\end{eqnarray} 
where $\lambda$ is a regularization parameter. Often \eqref{eq:logitL1} leads to a compact model with a `good' 
predictive accuracy. We will use the R package {\it glmnet} \cite{glmnet2010}. 
\subsection{Features}
\label{section:methodRepresentation}
Perhaps the most important dimension of a study is the feature, as it determines the classification 
problem for subsequent analysis. In our structured analysis model, the features are related to the feature 
error. A careful examination of the features can help gauge its strength, thus give insights on whether it is 
worthwhile to work further on feature extraction, or to improve the algorithm, or simply try 
to get a larger sample. It would also help in comparing two sets of features, or to give clues on 
the marginal benefit of a set of new features. Among our tools for examining the features are 
the covariance matrix, the distance of separation, and feature importance profiling etc. 
\\
\\
Our assessment of the features consists of an inspection on the covariance matrices, the computation 
of the distance of separation for relevant features, the generation of a feature importance profile, and, 
possibly, feature selection. As we have discussed the covariance matrix and
the distance of separation in Section~\ref{section:theory}, here we only discuss feature importance while 
omitting feature selection as it is too big a topic 
(readers can refer to \cite{LiuMotoda1998,GuyonElisseeff2003,TangAlelyaniLiu2014} and references therein). 
\\
\\
We recommend the use of RF to produce a feature importance profile. There are two feature importance metrics in RF, 
one based on the Gini index and the other permutation accuracy \cite{CART, RF}. We consider the later here, as it 
is often considered superior. The idea is as follows. Randomly permute the values of a feature, say, the $i^{th}$ feature, 
then its association with the response $Y$ is broken. When this feature, along with those un-permuted features, is 
used for prediction, the accuracy tends to decrease. The difference in the prediction accuracy before and after permuting 
the $i^{th}$ feature can then be used as a measure of its importance. 
\begin{figure}[h]
\centering
\begin{center}
\hspace{0cm}
\includegraphics[scale=0.64,clip]{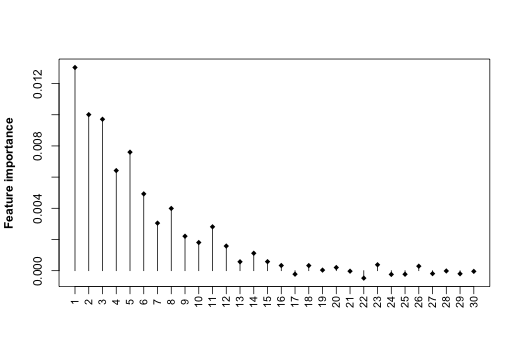}
\end{center}
\abovecaptionskip=-25pt
\caption{\it Feature importance by RF for the given example. } 
\label{figure:varImportanceEg}
\end{figure}
\\
\\
In the following, we use a two-component Gaussian mixture as an example to
demonstrate the use of RF for feature importance profiling. The Gaussian mixture is defined as in \eqref{eq:gm} with
\begin{equation*}
\bm{\mu}=(0.2,0.19,0.18,...,0.01,0,0,...0) \in \mathbb{R}^{30}, \bm{\Sigma}_{i,j}=0.1^{|i-j|}.
\end{equation*}
Thus the importance of features decreases with their feature index, with the last 10 features being purely noise features.
Figure~\ref{figure:varImportanceEg} shows the importance of features, ordered by their indices. It can be seen the feature
importance as produced by RF agrees fairly well with the generating model.
\section{Results}
\label{section:results}
In our experiments, we study the land-use classification using a study site in the Pearl River Delta 
region. Our experiments center around the four dimensions of a study. We explore a number of important 
aspects of land-use classification with structured analysis. This includes a study on the 
predictive accuracy of a classifier with varying sample sizes, marginal benefits of spectral, textural or 
location features, feature importance profiling, which land-use types are more difficult to classify than 
others, and predictive performance under `small' sample size. This is different from usual studies in remote 
sensing which usually focus on the prediction accuracy. 
\begin{table}[h]
\begin{center}
\begin{tabular}{rr|l|l} \toprule    
&\textbf{Experiments}                                &\textbf{Results}                                       & \textbf{Dimension of relevance}  \\
    \midrule
&{\it Sample size and performance}          & Figure~\ref{figure:errorSample}             & Sample, algorithm, features \\
&{\it Label noise}                                       & Figure~\ref{figure:errCont}                      & Error, sample, algorithm \\
&{\it Small sample performance}               & Figure~\ref{figure:errorSampleSmall}     & Sample, algorithm, features \\
&{\it Marginal benefits}                               &  Figure~\ref{figure:PairSeparation}        & Features \\
&                                                                 & Figure~\ref{figure:errorSample}             & Features\\
& {\it Distance of separation}                      & Figure~\ref{figure:PairSeparation}         & Features \\
&  {\it Covariance matrix}                            & Figure~\ref{figure:mapCovariance}       & Features \\
& {\it Feature importance}                           & Figure~\ref{figure:varImportance}         & Features, algorithm \\
& {\it Confusion matrix}                               & Table~\ref{table:confusion}                    & Error, algorithm, features \\
&  {\it Difficult land-types}                            & Figure~\ref{figure:orchardForest}          & Error, sample, features \\                                                              
&                                                                 & Figure~\ref{figure:industryAll}                & Error, sample, features \\
&                                                                 & Figure~\ref{figure:industryDownsample} &Error, sample, algorithm\\
    \bottomrule
\end{tabular}
\end{center}
\caption{\it A summary of experiments and their relevance to our structure analysis.}
\label{table:experiments}
\end{table}
\normalsize
\\
\\
Table~\ref{table:experiments} summarizes the experiments we conduct, and their relevance to the four
dimensions of a study. It should be noted that any experimental result is related to all the four dimensions, 
the table lists those dimensions that we view as the most relevant. Also note that, as many studies 
have mostly dealt with the algorithms dimension, we focus less on the algorithms in our study. For the rest of 
this section, we present details of our experiments and results.
\subsection{Sample size and performance}
Labeling in remote sensing studies is expensive, as it requires a verification to ground truth 
for which often a field trip is required. It is important to assess the effect of the sample size to the error rate. 
We explore the predictive accuracy of RF and $L_1$ logistic regression with different sample sizes.
\begin{figure}[h]
\centering
\begin{center}
\hspace{0cm}
\includegraphics[scale=0.35,clip]{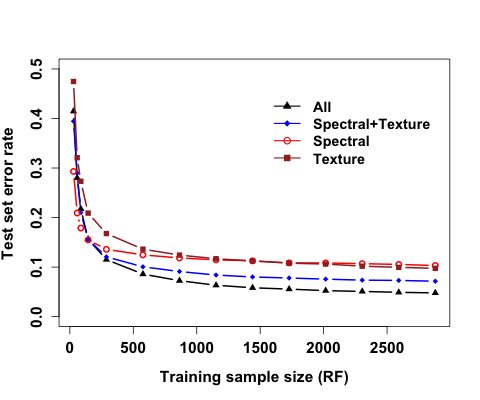}
\includegraphics[scale=0.35,clip]{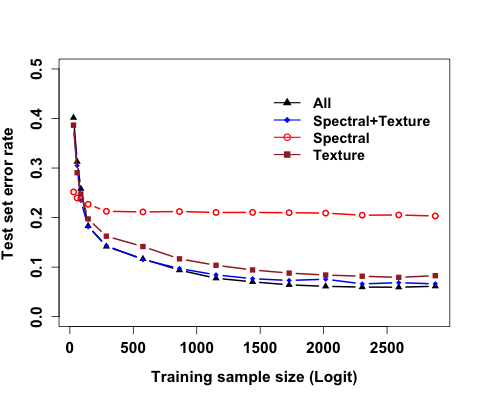}
\end{center}
\abovecaptionskip=-5pt
\caption{\it Error rates for spectral features, texture features, spectral and texture combined, and with 
additional location features when the sample size varies.  The two panels correspond to RF and $L_1$ 
logistic regression, respectively. `Logit' is short for logistic regression.} 
\label{figure:errorSample}
\end{figure}
\noindent
Figure~\ref{figure:errorSample} shows the error rates for varying sample sizes when using 4 different 
sets of features, including: 1) spectral features alone; 2) texture features alone; 3) combination 
of the two; 4) the combination with additional location features (latitude and longitude). In all 4 cases, 
there is an overall decreasing trend in the error rates when increasing the sample size. The two plots 
show very similar patterns except that the error rate curve with spectral features alone quickly levels 
off for $L_1$ logistic regression. 
\\
\\
This implies that even if further enlarging the training sample, there would still be a gap between the 
empirical and the Bayes error when using only the spectral features (totally 6). Clearly a logistic 
regression model, as a linear model, would converge very quickly on 6 variables. Thus such 
a gap is likely caused by the fact that the richness of the logistic regression models (using only the 6 spectral 
features) is not sufficient to match the complexity of the problem thus a non-vanishing approximation error.
\\
\\
Figure~\ref{figure:errorSample} suggests that, in all cases, further increasing the sample size may not gain 
much in reducing the 
overall error. This is bacause the convergence error is close to 0 since the curves already level off. As many different 
algorithms have been tried \cite{LiWWHG2014}, the approximation error should be very small for the best performing 
algorithm (see also the discussion on results by confusion matrix in Section~\ref{section:hardLand}). Thus it may be 
more worthwhile to explore the features dimension than the algorithms dimension. This is an insight we arrive at by 
exploring the sample and the algorithms dimensions. Later in Section~\ref{section:hardLand}, we will give clues on 
what kind of new features are likely worthwhile to further explore.          
\subsection{Error rates under label noise}
As mentioned in Section~\ref{section:methodData}, we will evaluate error rates under varying degrees of label noises.
An $\epsilon$ proportion of the training sample is randomly selected, and then their labels are flipped uniformly at random 
to a different label. The resulting sample is a contaminated version of the original one. The classifier will then be trained on 
the contaminated sample, and predictive accuracy evaluated on the clean (uncontaminated) test sample. 
\begin{figure}[h]
\vspace{-0.16in}
\centering
\begin{center}
\hspace{0cm}
\includegraphics[scale=0.52,clip]{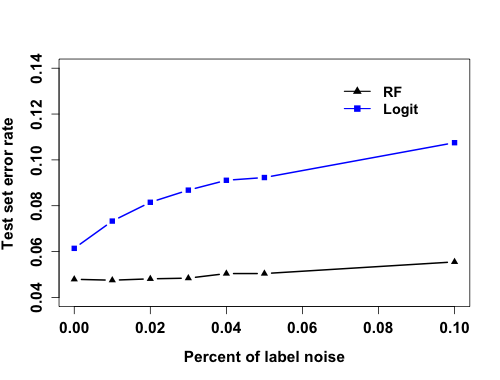}
\end{center}
\abovecaptionskip=-5pt
\caption{\it Error rates under label noises. The proportion of training sample with label noise is taken from 
\{0.01, 0.02, 0.03, 0.04, 0.05, 0.10\}. } 
\label{figure:errCont}
\end{figure}
\\
\\
Figure~\ref{figure:errCont} shows error rates as $\epsilon$ varies over the set $\{0.01, 0.02, 0.03, 0.04, 0.05, 0.10\}$. 
The error rate for $L_1$ logistic regression increases notably as 
$\epsilon$ increases from 0.01 to 0.10. However, the error curve for RF remains fairly flat. This indicates that
RF is more resistant to label noise than $L_1$ logistic regression. The almost flat trend of the error curve for 
RF would allow us to confidently conclude that either the original label noise (extrapolated from the curve) is 
very small or its impact is negligible when using RF.    
\subsection{Marginal benefit of spectral, texture and location features}
\begin{figure}[h]
\vspace{-0.16in}
\centering
\begin{center}
\hspace{0cm}
\includegraphics[scale=0.62,clip]{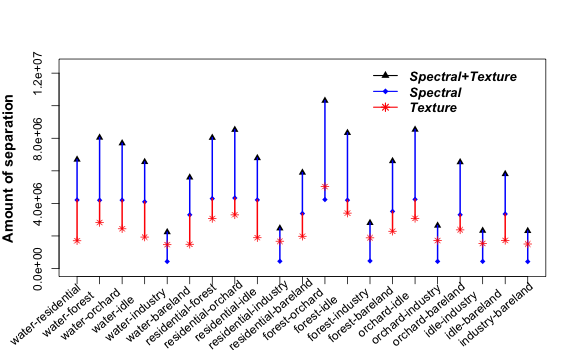}
\end{center}
\caption{\it Distance of separation of all pairs of land types. } 
\label{figure:PairSeparation}
\end{figure}

\begin{figure}[h]
\vspace{-0.1in}
\centering
\begin{center}
\hspace{0cm}
\includegraphics[scale=0.42,clip]{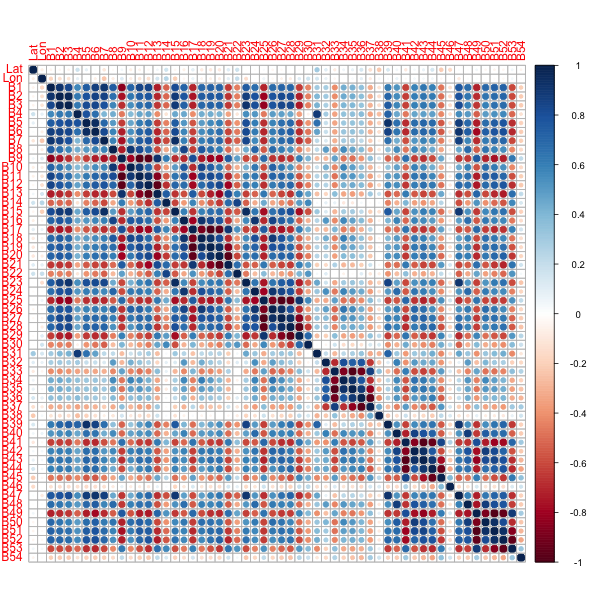}
\end{center}
\abovecaptionskip=-5pt
\caption{\it Heatmap of the covariance matrix. } 
\label{figure:mapCovariance}
\end{figure}

Many existing studies suggest that considering multiple features from different domains may be helpful to 
land classification \cite{ZhangZhangZhangTao2015, ZhangZhangDuTao2018}. So it is natural to expect that combing 
the texture features and spectral features would do better than using either alone. 
As shown in Figure~\ref{figure:errorSample}, this is the case when the sample size is not `small'. Here, by `small' 
we mean a sample size less than about 20 instances per land-use type. Such a cutoff is consistent 
with recommendations made in \cite{LiWWHG2014}. This could be explained by the {\it distance of separation}. 
\\
\\
Figure~\ref{figure:PairSeparation} shows the distance of separation for all pairs of land-use types (the distance 
of separation is only defined for a two-class classification problem in this work, and extension to multiple-class 
will be studied in our future work). It can be seen that for all pairs of land-use types, the distance of separation 
increases substantially when combining the spectral and texture features. Thus, as long as the training sample 
is large enough and the family of classifiers is rich enough, we will see reduced empirical error rates. 
\\
\\
However, the improvement with RF when combining spectral
and texture features (or adding spectral features in $L_1$ logistic regression when there are already 
texture features), which is about 2-3\%, is far less than expected.
In other words, the marginal benefits of spectral or texture features are small w.r.t. the other. This again 
could be understood from Figure~\ref{figure:PairSeparation} where the distance of separation 
between any pair is already big using either the spectral features or the texture features alone. A big distance of 
separation implies a small Bayes error. Thus, the room for improvement is small, and the marginal benefit 
of either the texture features or the spectral features w.r.t. the other is small. 
\\
\\
Another observation from Figure~\ref{figure:errorSample} is that the spectral features and the texture 
features lead to similar empirical error rates (for RF only) when the sample size is not small. This can 
also be explained by Figure~\ref{figure:PairSeparation} where all the distances of separation for the 
spectral and texture features are large and of a similar magnitude thus the Bayes error for both cases 
would be close to 0. Regarding the Bayes error of this land classification problem with all features, 
we expect it be fairly close to 0 (the error rate when using RF is about 4.79\%). The empirical error we get 
is a little upper-biased, due possibly to the discrepancy in class distribution between the test and the 
training sample according to discussions in Section~\ref{section:siteData}. This is further confirmed 
by Figure~\ref{figure:industryDownsample}.
\\
\\
In Figure~\ref{figure:errorSample}, we also observe that adding location features (i.e., latitude and longitude) 
leads to noticeable reduction in the error rates when the sample size is large. In the case of RF, the error rate 
reduces by about 2\%. This may be a little surprising, but could be understood by Theorem~\ref{thm:largerSeparation}.
A low `dependence' between the location and other features can be seen clearly in 
Figure~\ref{figure:mapCovariance} ({\it indicated by the bright colour in the first two rows and columns}). Thus, by 
Theorem~\ref{thm:largerSeparation}, we would expect an increased distance of separation when adding the 
location features thus a lower Bayes error. With a large sample size, this translates to a reduced 
error rate. In other words, the location features have a positive marginal benefit w.r.t. the spectral 
and texture features. Note that we cannot use Theorem~\ref{thm:largerSeparation} to explain the positive 
marginal benefits of either the spectral or the texture features w.r.t. the other, as clearly these features are 
highly correlated by Figure~\ref{figure:mapCovariance} (indicated by dark colour).   
\subsection{Performance under small sample sizes}
\label{section:smallSample}
\begin{figure}[h]
\begin{center}
\hspace{0cm}
\includegraphics[scale=0.35,clip]{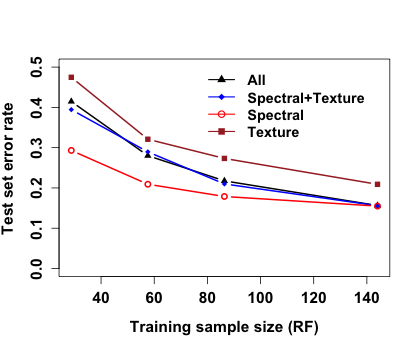}
\includegraphics[scale=0.35,clip]{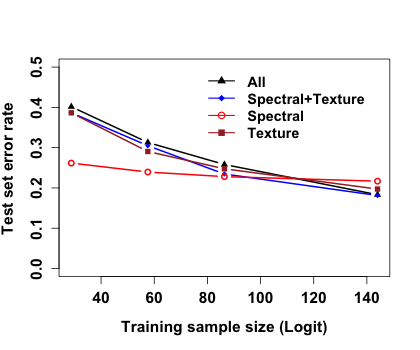}
\end{center}
\abovecaptionskip=-5pt
\caption{\it Error rates for spectral features, texture features, spectral and texture combined, 
and with additional location features under small  sample sizes. The two panels correspond 
to RF and logistic regression, respectively. } 
\label{figure:errorSampleSmall}
\end{figure}

As mentioned before, it is often not feasible to obtain a large sample, especially for a new study.
Here, we explore small training sample with sizes ranging from 28 to 140, or about 4 to 20 observations 
per land-use type. We observe that in such cases, 
combining spectral and texture features is no longer beneficial. Instead using spectral features alone
actually outperforms the combination of both. This can be seen from Figure~\ref{figure:errorSampleSmall} 
which is a closeup-view of Figure~\ref{figure:errorSample}. For such small sample sizes, combining 
the spectral and texture features would increase the data dimension to 54 (6 spectral features and 48 
texture features) thus the {\it curse of dimensionality} \cite{HTF2001} phenomenon occurs and the 
performance of the classifier deteriorates. This reaffirms that the usual recommendation \cite{LiWWHG2014} 
of having at least 20 observations per land-use type is reasonable. 
\\
\\
Additionally, we observe that, when the sample is `small', the spectral features are 
more `efficient' than the texture features. This is again the result of sample size and model richness 
tradeoff as the number of spectral features and texture features are 6 and 48, respectively thus much 
smaller convergence error while not substantially higher approximation error when using spectral features 
alone. Also, Figure~\ref{figure:PairSeparation} shows that, in many cases (almost all but those involving 
the `Industry' class), the spectral features have a larger distance of separation (thus smaller feature 
error). We view this as an implication of the strength of the spectral features.  
\subsection{Important features}
\label{section:impFeatures}
\begin{figure}[h]
\centering
\begin{center}
\vspace{-10pt}
\hspace{0cm}
\includegraphics[scale=0.64,clip]{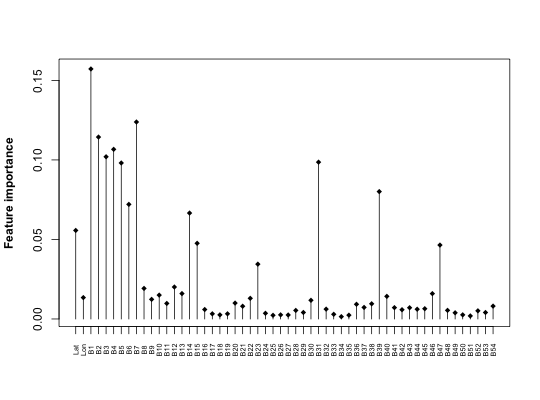}
\end{center}
\abovecaptionskip=-18pt
\caption{\it Variable importance produced by RF. } 
\label{figure:varImportance}
\end{figure}

Figure~\ref{figure:errorSample} and Figure~\ref{figure:errorSampleSmall} suggest that when the training 
sample is small, feature selection may be desirable during model fitting. 
Both RF and $L_1$ logistic regression have feature selection capability. In the following, 
we only report results obtained by RF as it has a built-in tool in producing feature importance.
\\
\\
Figure~\ref{figure:varImportance} shows the feature importance profile by RF. The top 10 most important features 
are B1-7, B31, B39, B14. Surprisingly, there is a major overlap with the spectral features B1-6. But this is 
consistent with our previous statements (c.f. Section~\ref{section:smallSample}) that the spectral features 
are `strong' features. Additional important features are B7, B31, B39, B14, which---except B14 the correlation 
texture feature for TM band 1---are the mean of texture feature values for TM band 1, 4, and 5, respectively
(the next three important features are B15, B47 and B23, which are the mean of texture features for TM 
band 2, 6 and 3, respectively). This is because the mean values carry a lot of information. 
Additionally, `Latitude' is an important feature (and `Longitude', which is more important 
than many texture features). This makes sense, as there is a high correlation between neighboring pixels in the 
image, and the land-use information of an image pixel has a strong predictive ability about that of its neighbors.       
\subsection{Which land-use types are harder to classify?}
\label{section:hardLand}
\begin{figure}[h]
\centering
\begin{center}
\hspace{0cm}
\includegraphics[scale=0.5,clip]{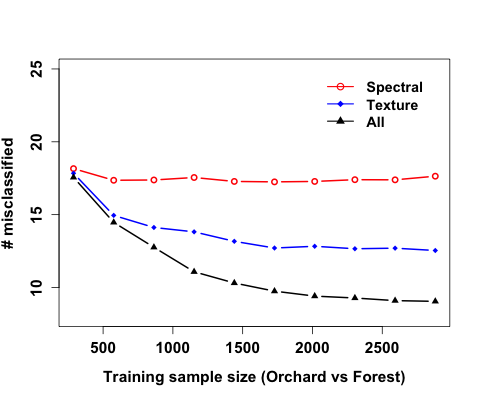}
\end{center}
\abovecaptionskip=-5pt
\caption{\it Number of misclassified instances in distinguishing `Orchard' from `Forest' as sample size increases. The
three curve correspond to using spectral feature, texture features, and the two combined, respectively.} 
\label{figure:orchardForest}
\end{figure}

\begin{figure}[h]
\centering
\begin{center}
\hspace{0cm}
\includegraphics[scale=0.5,clip]{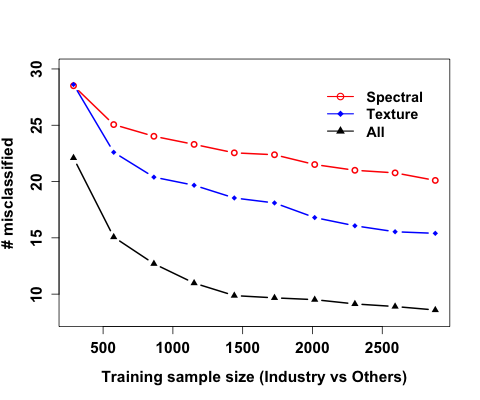}
\end{center}
\abovecaptionskip=-5pt
\caption{\it Number of misclassified instances in distinguishing `Industry' from other land-use types as sample size increases. The
three curve correspond to using spectral feature, texture features, and the two combined, respectively.} 
\label{figure:industryAll}
\end{figure}

\begin{figure}[h]
\centering
\begin{center}
\hspace{0cm}
\includegraphics[scale=0.5,clip]{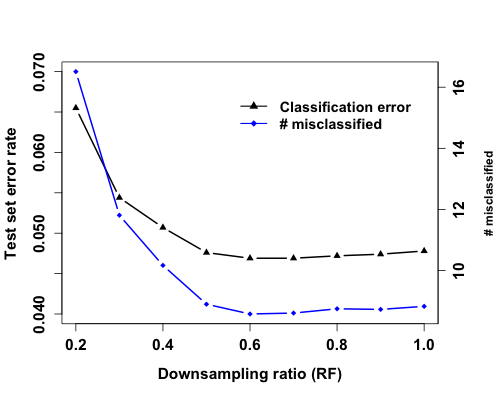}
\end{center}
\abovecaptionskip=-5pt
\caption{\it Error rate and the number of misclassified instances involving `Industry' when downsampling the 
`Industry' class in the training sample. } 
\label{figure:industryDownsample}
\end{figure}

As a matter of fact, the classification of different land-use types involves a varying level of difficulty. 
It will be helpful to figure out those land-use types that are harder than others, and a further study of which
will likely reduce the overall error rate. The confusion matrix is an ideal tool for this purpose. We will
use the confusion matrix produced by RF for the sake of convenience (logistic regression has comparable 
predictive accuracy as RF when using the texture features or all the features, but inferior with
spectral features alone). 
\\
\\
Table~\ref{table:confusion} shows the confusion matrices. There are three numbers
in each cell, indicating results produced with all features, spectral features only, and texture features only, 
respectively. We have several observations
\begin{itemize}
\item Between many pairs of land-use types, there is a zero classification error. This can be explained by their 
large distance of separation as shown in Figure~\ref{figure:PairSeparation} and that the training sample size 
(which is 2880) is sufficiently large (error curves level off in Figure~\ref{figure:errorSample}) thus small convergence 
errors. We hypothesize that the approximation error (not observable from experiments) resulting from using 
the RF classifiers is also `small'.
\item The major misclassification occurs between two cases, `Forest'  vs Orchard', and `Residential' vs `Industry'. 
These two cases may be of a different nature though; see detailed discussions later. 
\item Overall, `Industry' is a land-use type that is difficult to classify. It is often mis-classified as other types, 
or other types classified as `Industry'. This may be explained by the relatively smaller distance of separation 
between `Industry' and other land-use types, as shown in Figure~\ref{figure:PairSeparation}.
\item Combining spectral and texture features helps the most in distinguishing `Forest' vs 'Orchard', and 
`Residential' vs `Industry'. This is also true for varying sample sizes and more pronounced for larger sample sizes, 
according to Figure~\ref{figure:orchardForest} and Figure~\ref{figure:industryAll}. This later observation is expected as 
the convergence error would be smaller for larger sample sizes.
\end{itemize}
\begin{table}[h]
\begin{center}
\resizebox{0.9\textwidth}{!}{%
\begin{tabular}{rr|c|c|c|c|c|c|c}    \toprule
&                       &\textbf{Bareland} &\textbf{Forest}  &\textbf{Idle}  &\textbf{Industry}  &\textbf{Orchard}  &\textbf{Residential}  &\textbf{Water}  \\
    \midrule
&\textbf{Bareland}          &42           &0         &0             &2            &0           &0         &0  \\
&				  &42/41     &0/0        &0/0          &2/2         &0/0        &0/1      &0/0 \\ \cmidrule{1-9}
&\textbf{Forest}               &0            &73         &1            &0            &9            &0         &0  \\
&                        &0/0         &63/69    &3/0         &0/0          &16/13    &0/0      &1/1  \\ \cmidrule{1-9}
&\textbf{Idle}                  &0            &1          &44           &0             &0            &0         &0     \\
&                        &0/1         &0/1       &44/41      &0/0          &1/0         &0/2      &0/0  \\ \cmidrule{1-9}
&\textbf{Industry}           &2            &0          &1             &66           &0            &2         &0  \\
&                       &2/3          &0/0       &1/2          &65/64      &0/0         &3/2      &0/0\\ \cmidrule{1-9}
&\textbf{Orchard}           &0            &0           &0            &0             &48           &0         &0  \\
&                       &0/0         &2/0        &0/2         &0/0          &46/45      &0/0      &0/1\\ \cmidrule{1-9}
&\textbf{Residential}       &0            &0          &0             &1             &0            &89        &1  \\
&                       &0/0         &0/0        &0/2          &11/5        &0/0         &79/82   &1/2 \\ \cmidrule{1-9}
&\textbf{Water}              &0            &0           &0             &0             &0             &0        &41  \\
&                       &0/0         &0/0        &0/0          &0/0          &0/0          &0/0     &41/41 \\
    \bottomrule
\end{tabular}}
\end{center}
\caption{\it Confusion matrix produced by RF. In each cell, there are three numbers for
results using all features, spectral features or texture features alone, respectively.}
\label{table:confusion}
\end{table}
Figure~\ref{figure:PairSeparation} shows a relatively small distance of separation for `Residential' vs 
`Industry'. This implies that it may be more productive to add additional informative features to reduce 
the feature error. The later is consistent with our understanding---the difficulty in distinguishing `Residential' 
vs `Industry' lies in the fact that both land-use types are highly heterogeneous, and these two are similar 
in many aspects. 
\\
\\
How to further reduce the error in `Residential' vs `Industry'? One idea is to 
over-represent the `Industry' class in the training sample as in \cite{LiWWHG2014, AdaBoost}. In the 
training sample, the land type `Industry' already has 960 instances, much larger than other land types so it is 
already over-represented. To see the effect of over-representation, as we do not have a larger sample, we downsample 
the `Industry' class and use the trend to infer the effect of overrepresentation. 
\\
\\
Figure~\ref{figure:industryDownsample}
shows the error rate and the number of misclassified test instances involving `Industry' (i.e., all instances with a 
land-use type `Industry' or been classified as `Industry') with different down-sampling ratios for the `Industry'
land-use type. It can be seen that the smallest error rate
for both are achieved at roughly a sampling ratio of 0.6 (to match the class distribution in the test sample, the sampling
ratio should be around 0.4). So overrepresentation helps a little bit, but further overrepresentation will not help. This is likely 
due to the fact that further increasing the sampling ratio would cause more discrepancy between the training and the 
test sample thus a larger error rate. As the convergence error is already small (level-off on various error curves) and the 
approximation error with RF is hypothesized to be small, a possible future direction is to try reducing the feature error, 
i.e., look for features that would better distinguish `Residential' vs `Industry'. 
\section{Conclusions}
\label{section:conclusions}
In this paper, we propose a structured approach for the analysis of a land-use classification study. Under our 
model for structured analysis, we view the outputs (e.g., error), one dimension of a study, as a result of the 
interplay of other three dimensions, including feature, sample, and algorithms. Moreover, the land-use classification 
error can be further decomposed into error components according to these three dimensions. Such a structural 
decomposition of entities involved in land-use classification would help us better understand the nature of a 
land classification study, and potentially allow better trace of the merits or difficulty of a study to a more concrete 
source entity, or a more refined characterization of the difficulty of the problem. The analysis of a remote sensing 
image about a study site in Guangzhou, China, is used to demonstrate how a structured analysis could be carried 
out. We are able to identify a few possible directions for future studies that would potentially further reduce the 
land-use classification error; such information are typically beyond a usual land-use classification study. We expect 
the structured analysis as we have proposed will inform practices in the analysis of remote sensing images, and 
help advance the state-of-the-art of study on the land-use classification problem. 
\section{Proof of Theorem~\ref{thm:largerSeparation}} 
\label{section:appendix}
\begin{proof}
The proof uses perturbation analysis \cite{HuangYanNIPS2008, YanHuangJordan2009, TACOMA}, and follows 
a similar line of arguments as \cite{TACOMA}.
\\
\\
To simplify notations, write \[\bm{A} = \left[
\begin{array}{cc}
\bm{A}_{11}              & \bm{0}\\
\bm{0}            & \bm{A}_{22}\\
\end{array}
\right]\] and \[\bm{E} = \left[
\begin{array}{cc}
\bm{0}              & \bm{E}_{12}\\
\bm{E}_{12}^T             & \bm{0}\\
\end{array}
\right]\] where $\bm{0}$'s denote null matrices with appropriate
dimensions and $\bm{E}_{12}=\bm{A}_{12}$. Thus $\bm{\Sigma} =\bm{A}+\bm{E}$. To facilitate
the calculation of
$\bm{u}_{\bm{\mathcal{F}}}^T\bm{\Sigma}^{-1}\bm{u}_{\bm{\mathcal{F}}}$, we will
first derive a Taylor's series expansion result. Since
\begin{equation}
||\bm{E}||_F^2=2||\bm{E}_{12}||_F^2=2||\bm{A}_{12}||_F^2
\end{equation}
and $||\bm{A}^{-1}\bm{E}||_F \leq ||\bm{A}^{-1}||_F ||\bm{E}||_F \leq
\sqrt{\mbox{rank}(\bm{A}^{-1})}||\bm{A}^{-1}||_2||\bm{E}||_F$ (by the sub-multiplicative property of the Frobenius 
norm \cite{GolubVanLoan1989} and the boundedness of the eigenvalues of $\bm{A}^{-1}$), 
the following Taylor's series expansion is valid
\begin{equation}
(\bm{I}+\bm{A}^{-1}\bm{E})^{-1}=\bm{I} - \bm{A}^{-1}\bm{E} +O[(\bm{A}^{-1}\bm{E})^2],
\end{equation}
where $O$ in the above is the big-O notation. It follows that
\begin{eqnarray}
\bm{u}_{\bm{\mathcal{F}}}^T \bm{\Sigma}^{-1} \bm{u}_{\bm{\mathcal{F}}} &=&
[\bm{u}_1, \bm{u}_2]^T (\bm{A}+\bm{E})^{-1}
[\bm{u}_1, \bm{u}_2] \nonumber\\
&=& [\bm{u}_1, \bm{u}_2]^T (I - \bm{A}^{-1}\bm{E} +O[(\bm{A}^{-1}\bm{E})^2])\bm{A}^{-1} [\bm{u}_1,
\bm{u}_2] \nonumber\\
&=& [\bm{u}_1, \bm{u}_2]^T (\bm{A}^{-1}-\bm{A}^{-1}\bm{E}\bm{A}^{-1} + O[(\bm{A}^{-1}\bm{E})^2] \bm{A}^{-1}) [\bm{u}_1,
\bm{u}_2] \nonumber\\
&=& [\bm{u}_1, \bm{u}_2]^T \bm{A}^{-1} [\bm{u}_1, \bm{u}_2] - [\bm{u}_1,
\bm{u}_2]^T \bm{A}^{-1}\bm{E}\bm{A}^{-1} [\bm{u}_1, \bm{u}_2] \nonumber\\
&& +[\bm{u}_1, \bm{u}_2]^T
O[(\bm{A}^{-1}\bm{E})^2] \bm{A}^{-1} [\bm{u}_1,
\bm{u}_2] \nonumber\\
&=& \mathcal{S}_1 + \mathcal{S}_2 + \mathcal{S}_3.
\end{eqnarray}
$\mathcal{S}_1$ can be calculated as
\begin{equation}
\label{eq:quadEq} \mathcal{S}_1 = [\bm{u}_1, \bm{u}_2]^T\left[
\begin{array}{cc}
\bm{A}_{11}              & \bm{0}\\
\bm{0}            & \bm{A}_{22}\\
\end{array}
\right]^{-1}[\bm{u}_1, \bm{u}_2] = \bm{u}_1^T \bm{A}_{11}^{-1} \bm{u}_1 +
\bm{u}_2^T \bm{A}_{22}^{-1} \bm{u}_2.
\end{equation}
We have
\begin{equation}
\label{eq:normIneq1} ||\mathcal{S}_2 ||_F \leq
||\bm{u}||_F^2.||\bm{A}^{-1}||_F^2 ||\bm{E}||_F \leq C^2 ||\bm{A}^{-1}||_F^2 ||\bm{E}||_F =o(1)
\end{equation}
since $||\bm{A}||_F \leq \sqrt{\mbox{rank}(\bm{A})} ||\bm{A}||_2$ and $||\bm{A}^{-1}||_F \leq
\sqrt{\mbox{rank}(\bm{A}^{-1})} ||\bm{A}^{-1}||_2$ (where $||.||_2$ denote the 2-norm
of a matrix) and by the boundedness of the eigenvalues of $\bm{A}$ and
$\bm{A}^{-1}$. Similarly, we have
\begin{equation}
\label{eq:normIneq2} ||\mathcal{S}_3 ||_F \leq
||\bm{u}||_F^2 ||\bm{A}^{-1}||_F^3 ||\bm{E}||_F^2 =o(1).
\end{equation}
Combining equations \eqref{eq:quadEq}, \eqref{eq:normIneq1} and
\eqref{eq:normIneq2} yields
\begin{equation}
\bm{u}_{\bm{\mathcal{F}}}^T \bm{\Sigma}^{-1} \bm{u}_{\bm{\mathcal{F}}} = \bm{u}_1^T
\bm{A}_{11}^{-1} \bm{u}_1 + \bm{u}_2^T \bm{A}_{22}^{-1} \bm{u}_2 +o(1).
\end{equation}
Since both $\bm{u}_1^T \bm{A}_{11}^{-1} \bm{u}_1$ and $\bm{u}_2^T \bm{A}_{22}^{-1} \bm{u}_2$ are nonnegative, 
the claim of the theorem has been proved.
\end{proof}

\end{document}